\documentclass{article}

\usepackage[final]{neurips_2021_ml4ps}
\usepackage{graphicx}
\usepackage{natbib}
\setcitestyle{square,sort,comma,numbers}

\usepackage[utf8]{inputenc} 
\usepackage[T1]{fontenc}    
\usepackage{hyperref}       
\usepackage{url}            
\usepackage{booktabs}       
\usepackage{amsfonts}       
\usepackage{nicefrac}       
\usepackage{microtype}      
\usepackage{xcolor}         

\title{Weight Pruning and Uncertainty in Radio Galaxy Classification}

\author{%
  Devina Mohan \\
  Department of Physics \& Astronomy\\
  University of Manchester, UK\\
  \texttt{devina.mohan@postgrad.manchester.ac.uk} \\
  \And
  Anna Scaife\thanks{The Alan Turing Institute, 96 Euston Rd, London, UK \texttt{a.scaife@turing.ac.uk}} \\
  Department of Physics \& Astronomy\\
  University of Manchester, UK\\
  \texttt{anna.scaife@manchester.ac.uk} \\
}

\begin{document}

\maketitle

\begin{abstract}
 In this work we use variational inference to quantify the degree of epistemic uncertainty in model predictions of radio galaxy classification and show that the level of model posterior variance for individual test samples is correlated with human uncertainty when labelling radio galaxies. We explore the model performance and uncertainty calibration for a variety of different weight priors and suggest that a sparse prior produces more well-calibrated uncertainty estimates. Using the posterior distributions for individual weights, we show that signal-to-noise ratio (SNR) ranking allows pruning of the fully-connected layers to the level of 30\% without significant loss of performance, and that this pruning increases the predictive uncertainty in the model. Finally we show that, like other work in this field, we experience a cold posterior effect. We examine whether adapting the cost function in our model to accommodate model misspecification can compensate for this effect, but find that it does not make a significant difference. We also examine the effect of principled data augmentation and find that it improves upon the baseline but does not compensate for the observed effect fully. We interpret this as the cold posterior effect being due to 
 the overly effective curation of our training sample leading to likelihood misspecification, and raise this as a potential issue for Bayesian deep learning approaches to radio galaxy classification in future. 
\end{abstract}

\section{Introduction}
\label{sec:intro}

A new generation of radio astronomy facilities around the world such as the Low-Frequency Array  \citep[LOFAR;][]{VanHaarlem2013}, the Murchison Widefield Array \citep[MWA;][]{Beardsley2019}, the MeerKAT telescope \citep{Jarvis2016}, and the Australian SKA Pathfinder (ASKAP) telescope \citep{Johnston2008} are generating increasingly larger and larger data rates. In order to extract scientific impact from these facilities on reasonable timescales, a natural solution has been to automate the data processing as far as possible and this has lead to the increased adoption of machine learning methodologies. 

In particular for new sky surveys, automated classification algorithms are being developed to replace the \emph{by eye} approaches that were possible historically. In radio astronomy specifically, studies looking at morphological classification using convolutional neural networks (CNNs) and deep learning have become increasingly common, in particular with respect to the classification of radio galaxies. 

The Fanaroff-Riley (FR) classification of radio galaxies was introduced over four decades ago \citep{fr1974}, and has since been widely adopted and applied to many catalogues since then. The morphological divide seen in this classification scheme has historically been explained primarily as a consequence of differing jet dynamics. Fanaroff-Riley type I (FRI) radio galaxies have jets that are disrupted at shorter distances from the central super-massive black hole host and are therefore centrally brightened, whilst Fanaroff-Riley type II (FRII) radio galaxies have jets that remain relativistic to large distances, resulting in bright termination shocks. These observed structural differences may be due to the intrinsic power in the jets, but will also be influenced by local environmental densities \citep{bicknell1995,kaiserbest}.   

Intrinsic and environmental effects are difficult to disentangle using radio luminosity alone as systematic differences in particle content, environmental effects and radiative losses make radio luminosity an unreliable proxy for jet power \citep{croston2018}. Hence the use of morphology for inferring the environmental impact on radio galaxy populations is therefore important for gaining a better physical understanding of the FR dichotomy, and of the full morphological diversity of the population.

From a deep-learning perspective, the ground work for morphological classification in this field was done by \citep{aniyan2017}. This was followed by other works involving the use of deep learning in source classification \citep[e.g.][]{lukic2018,rgz,wu2018}. More recently, \citep{bowles2020} showed that an attention-gated CNN could perform classification of radio galaxies with equivalent performance to other applications in the literature, but using $\sim$50\% fewer learnable parameters, \citep{e2cnn} showed that using group-equivariant convolutional layers that preserved the rotational and reflectional isometries of the Euclidean group resulted in improved overall model performance and stability of model confidence for radio galaxies at different orientations, and \citep{bastien2020} generated synthetic populations of radio galaxies using structured variational inference.

With the exception of \citep{e2cnn}, there has been little work done on understanding the degree of confidence with which CNN models predict the class of individual radio galaxies; however, for radio astronomy, where modern astrophysical analysis is driven by population analyses, quantifying the confidence with which each object is assigned to a particular classification is crucial for understanding the propagation of uncertainties within that analysis. 

\vskip .1in
\noindent
\textbf{This work} In this work we use variational inference (VI) to quantify the degree of epistemic uncertainty in model predictions of radio galaxy classification. This differs from the approach of \citep{e2cnn} who used dropout as a Bayesian approximation \citep{gal2016} to estimate model confidence as a function of image orientation, which is just one specific aspect of model performance and not directly comparable to this work. We compare the variance of our posterior predictions to qualifications present in our test data that indicate the level of human confidence in assigning a classification label and show that model uncertainty is correlated with human uncertainty. 

An added advantage of Bayesian CNNs is their ability to identify the significance of individual weights within the model. In this work we show that weight signal-to-noise ratio (SNR) allows pruning of our fully-connected layers to the level of 30\% without significant loss of model performance. We also find that pruning based on a Fisher analysis is able to thin our model more effectively, to the level of 60\%, 
but that both pruning methods increase the uncertainty calibration error of the model in different ways.

Finally we show that, like other work in this field, we experience a \emph{cold posterior} effect \citep[see e.g.][]{wenzel2020}. As suggested by \citep{masegosa2019}, we examine whether a PAC Bayes approach adapting the cost function in our model to accommodate model misspecification can compensate for this effect, but find that it does not make a significant difference. Other works in this area have suggested that unprincipled data augmentation could be a contributing factor to the cold posterior effect \citep[e.g.][]{nabarro2021data, izmailov21}. We examine the effect of principled data augmentation and find that the cold posterior effect observed in our work \emph{reduces} slightly with data augmentation. Consequently we suggest that the cold posterior effect in this case is likely to be due to the manner in which the training sample is curated, rather than model misspecification. 


\section{Uncertainty in radio galaxy classification}
\label{sec:epistemic}

For this work we use the MiraBest radio galaxy dataset\footnote{The data used in this work is provided under a Creative Commons license  at \url{https://zenodo.org/record/4288837}}  \citep{porterzenodo}, which is based on the catalogue of \citep{mirabest2017}, who used a parent galaxy sample taken from \citep{bestheckman} that cross-matched the Sloan Digital Sky Survey \citep[SDSS;][]{sdss} data release 7 \citep[DR7;][]{sdssdr7} with the Northern VLA Sky Survey \citep[NVSS;][]{NVSS} and the Faint Images of the Radio Sky at Twenty centimetres \citep[FIRST;][]{FIRST}. The parameters of the dataset itself and the image preprocessing are described in \citep{e2cnn}. For this work we extract the objects labelled as FRI and FRII galaxies. This creates a binary dataset and we do not employ sub-classifications. Each object within the MiraBest dataset was assigned a confidence qualification: \emph{Confident} or \emph{Uncertain} by its original human classifiers. Training is performed on the \emph{Confident} subset.

\begin{figure*}
    \centering
    \includegraphics[width=\textwidth]{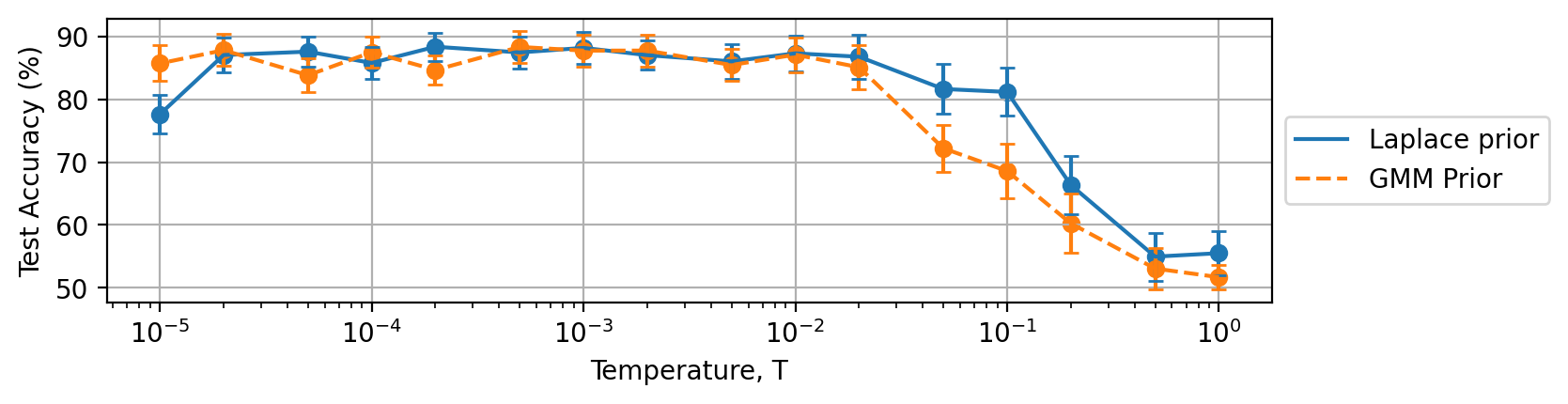}
    \caption{The “cold posterior” effect: for the MiraBest classification problem presented here we can improve the generalization performance significantly by cooling the posterior with a temperature $T\ll1$, deviating from the Bayes posterior. Data are shown for the BBB models with no data augmentation and the original ELBO cost function trained with a Laplace prior (solid blue line), and trained with a GMM prior (orange dashed line).}
    \label{fig:prior_coldpost}
\end{figure*}

We use an expanded LeNet-5 architecture incorporating two additional convolutional layers with 26 and 32 channels, respectively, and implement the variational inference approach described in \citep{blundell} using the Adam optimiser with an initial learning rate of $5 . 10^{-5}$. Hyper-parameters were tuned using a grid search. We use a Gaussian variational approximation to the true posterior and we consider four different prior distributions over the model parameters: a simple Gaussian, a two component Gaussian mixture model (GMM) prior, a Laplace prior and a Laplace Mixture Model (LMM) prior. We find that model performance is optimised using a Laplace prior. Using a GMM prior allows us to achieve comparable accuracy to the model trained with a Laplace prior but it has worse uncertainty calibration error and experiences a more significant cold posterior effect, see Figure~\ref{fig:prior_coldpost}. In the following analysis we report results for the Laplace prior.

We train the model in batches of 50 images and implement an early stopping criterion based on validation accuracy. We note that while other works have used deeper models for classification of radio galaxies, no significant difference in performance is seen for binary FR classification. We also note that, unlike previous works, we do not use any data augmentation in our training. Models were trained on an Intel i5 CPU for 500 epochs. Each training run took approximately $1$\,hour.

\begin{figure*}
    \centering
    \includegraphics[width=\textwidth]{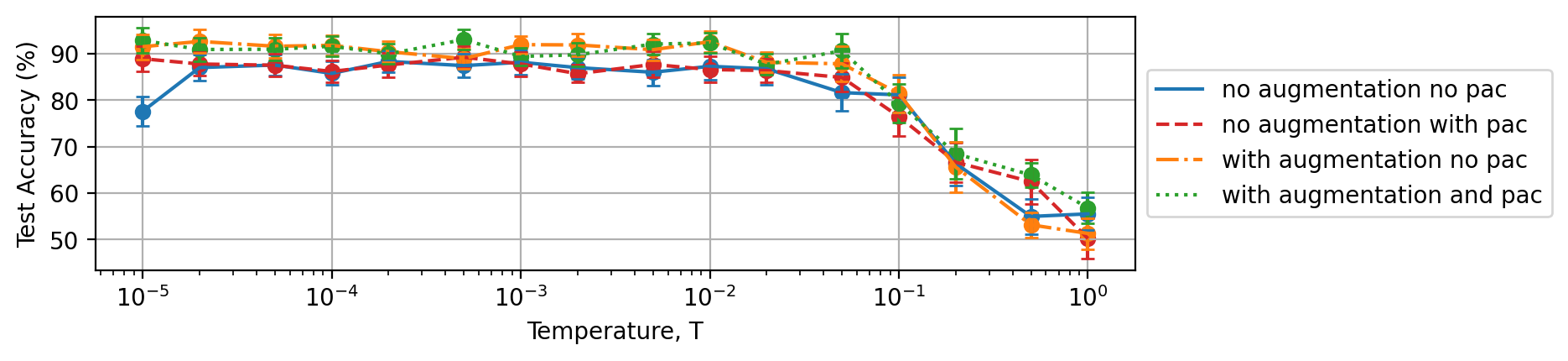}
    \caption{The “cold posterior” effect: Data are shown for the BBB model trained with a Laplace prior with no data augmentation and the original ELBO cost function (solid blue line), the BBB model with no data augmentation and the Masegosa posterior cost function (red dashed line), the BBB model with data augmentation and the original ELBO function (orange dot-dash line), and the BBB model with data augmentation and the Masegosa posterior cost function (green dotted line).}
    \label{fig:coldpost}
\end{figure*}

\vskip .1in
\noindent
\textbf{The cold posterior effect} It has been observed in the literature that in order to obtain good predictive performance from Bayesian neural networks, it is necessary to down-weight or \emph{lower the temperature} of the Bayesian posterior. Several explanations have been proposed to explain this effect, including model or prior misspecification \citep{wenzel2020}, and data augmentation or curation issues \citep{aitchison2021a}. We also observe this effect in the classification of radio galaxies in this work, see Figure~\ref{fig:coldpost}. 

Assuming that all models are likely to be misspecified at some level, \citep{masegosa2019} suggested that the use of more generalised posteriors that treat the true Bayes posterior as a solution of a loose PAC-Bayes generalization bound on the predictive cross-entropy might provide a more desirable approximation target than the Bayes posterior itself. We retrain our models using the PAC-Bayes cost function defined in \citep{masegosa2019}, but find that this does not make a significant difference to the observed cold posterior effect. 

If we include data augmentation using random rotations from 0 to 360\,degrees, the cold posterior effect observed in our work \emph{reduces} slightly, see Figure~\ref{fig:coldpost}. We suggest that this is because we have augmented the MiraBest dataset using principled methods that correspond to an informed prior for how radio galaxies are oriented, as radio galaxy class is assumed to be equivariant to orientation and chirality \citep[see e.g.][]{ntwaetsile2021,e2cnn}. Since the Masegosa posterior is a more complete test of model misspecification than the data augmentation used here is of likelihood misspecification, we suggest that a key element for exploring this problem in future may be the availability of radio astronomy training sets that do not only present average or consensus target labels, but instead include all individual labels from human classifiers. For the following results we use a posterior temperature of $T = 10^{-2}$.

\begin{figure}
\centerline{\includegraphics[width=0.3\textwidth]{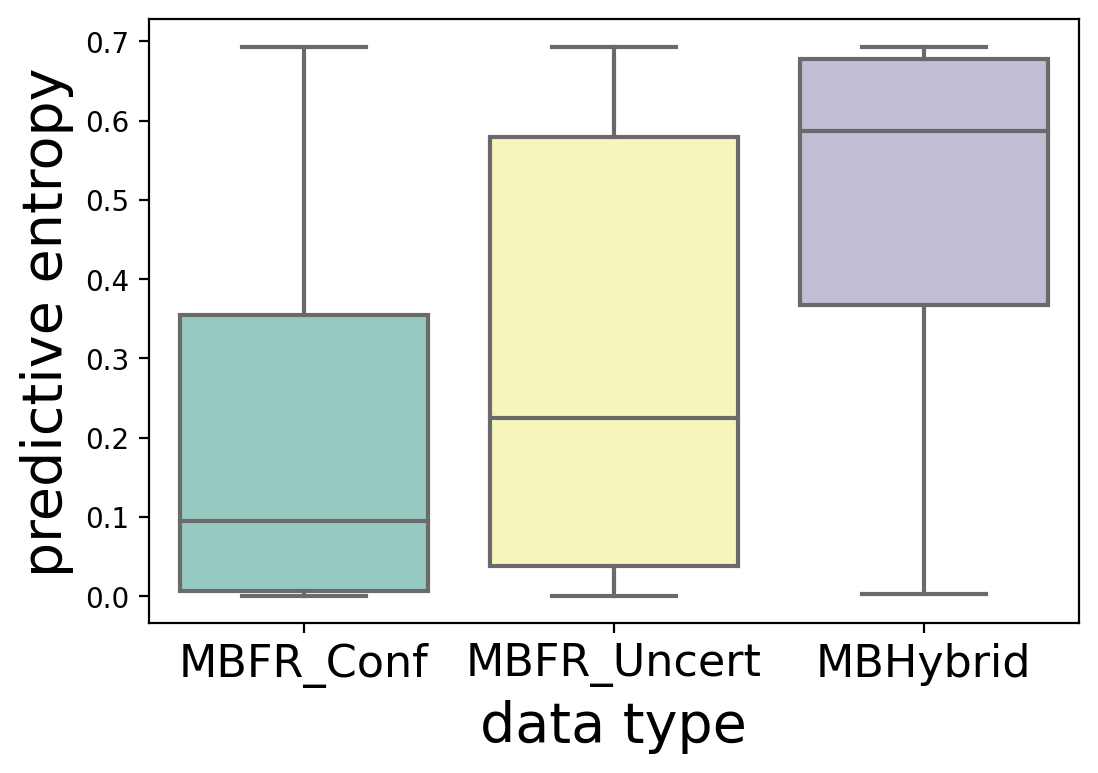}\qquad \includegraphics[width=0.3\textwidth]{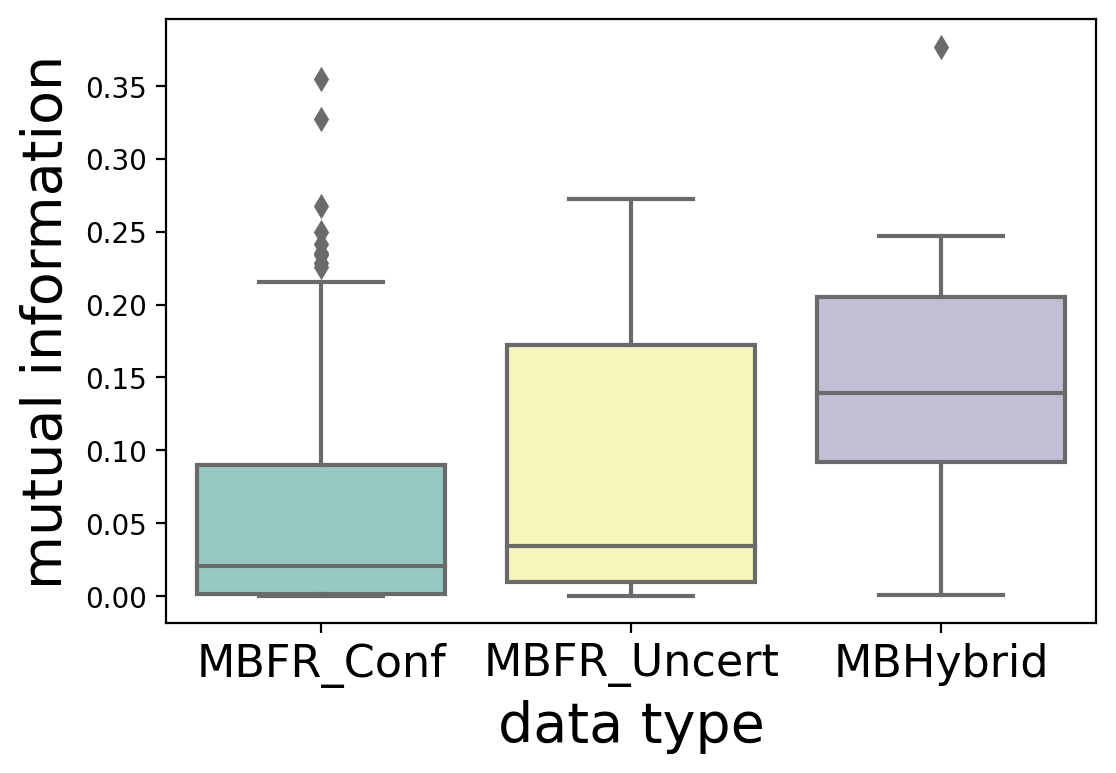}\qquad \includegraphics[width=0.3\textwidth]{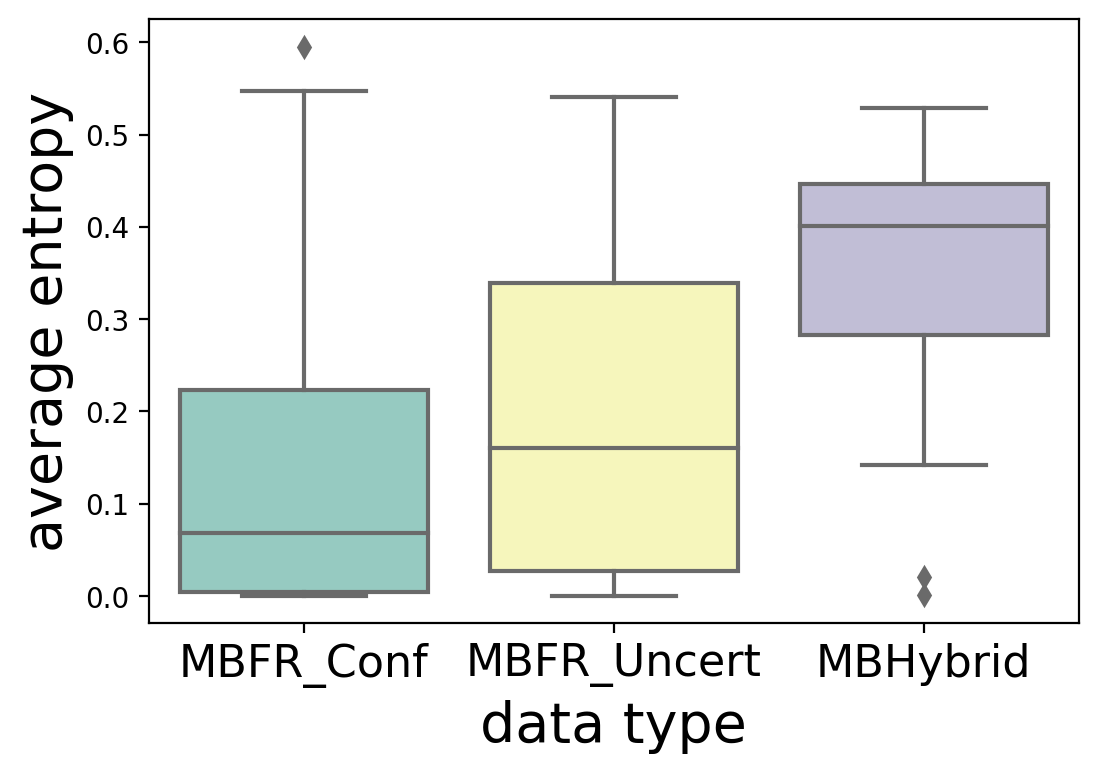}}

    \caption{Distribution of predictive uncertainty (entropy; left), epistemic uncertainty (mutual information; centre) and aleatoric uncertainty (average entropy; right) for the \emph{Confident} (MBFR\_Conf) and \emph{Uncertain} (MBFR\_Uncert) test samples. Data are also shown for the \emph{MiraBest} Hybrid test sample.}
    \label{fig:entropy}
\end{figure}

\vskip .1in
\noindent
\textbf{Model Confidence} As well as the MiraBest \emph{Confident} test set, we use 49 samples from the MiraBest \emph{Uncertain} subset to test the trained model's ability to correctly represent epistemic uncertainty. These samples can be considered to be drawn from the same data generating distribution as the MiraBest \emph{Confident} samples, but have a lower degree of belief in their classification. Using Monte Carlo samples obtained from the trained predictive posterior distribution, we obtain 200 Softmax probabilities for each test sample. Following \citep{gal2016}, we find that the predictive uncertainty (epistemic + aleatoric), as quantified by the entropy, and the model uncertainty (epistemic), as quantified by the mutual information of these samples, show that on average test data samples labelled as \emph{Uncertain} have higher epistemic uncertainty than those labelled as \emph{Confident}, see Figure~\ref{fig:entropy}. To quantify aleatoric uncertainty for in-distribution data samples for classical NNs \citep{mukhoti} demonstrate that the entropy of a single pass can be used. Here we extend that definition to our Bayesian NN and take the average entropy for a single input using MC samples to capture the aleatoric uncertainty associated with each data point.

We also examine the uncertainty metrics for the MiraBest \emph{Hybrid} subsample, which contains galaxies that have characteristics of both classes. 
We find that all uncertainty measures for this sample are even higher than for \emph{Uncertain} FRI and FRII type objects, see Figure~\ref{fig:entropy}. We also note that Confidently classified \emph{Hybrid} samples have higher predictive uncertainty than the Uncertain \emph{Hybrid} samples. This
could be because the Uncertain samples are more like the FRI/FRII galaxies that the model has seen during training, i.e. their classification as a \emph{Hybrid} was considered Uncertain by a human classifier because the morphology was biased towards one of the standard FRI or FRII classifications. In which case their predictive uncertainty might be expected to be lower since the model was trained to predict those morphologies.

\vskip .1in
\noindent
\textbf{Uncertainty calibration} We quantify the calibration of our posterior uncertainties using the class-wise expected Uncertainty Calibration Error (cUCE; \citep{cUCE}). We find that predictive entropy and average entropy for the unpruned model are better calibrated than the mutual information, see Table~\ref{tab:calibration}. 

\section{Weight pruning based on posterior variance}
\label{sec:pruning}

For a typical variational posterior such as the Gaussian distribution, the number of parameters in a Bayesian model will double compared its standard counterpart because both the mean and standard deviation values need to be learned. Consequently, to reduce the computational cost and memory overhead during deployment, network pruning approaches which remove uninformative parameters are often used. Several authors have also considered pruning to improve the generalisation performance of the network \citep[e.g.][]{lecun1990optimal}.

Following \citep{blundell}, we calculate the signal-to-noise ratio (SNR) of individual weights as $|\mu|/\sigma$, where the trained posterior distribution on a weight is $\mathcal{N}(\mu, \sigma^2)$. By ranking all of the model weights in order of SNR, we are able to prune a given percentage of the lowest SNR-valued weights. We then calculate the error on the test set using the average of 100 forward passes. 

The choice of prior directly influences the shape of the SNR distribution and the proportion of weights that can be pruned without affecting model performance on the test set. For the Laplace prior, we find that up to 30\% of the weights in the fully-connected layers can be pruned without a significant drop in performance. However, following the work of \citep{tu2016reducing}, we also explore a pruning approach that uses the Fisher Information Matrix (FIM) of the weights. As also observed by \citep{tu2016reducing}, pruning the weights based on the Fisher information alone does not allow for a large number of parameters to be pruned effectively because many values in the FIM diagonal are close to zero; however, combining Fisher-based pruning and magnitude-based pruning allows for a larger number of weights to be pruned - up to 60\% in this case.

We find that both the SNR and Fisher pruned models have higher predictive uncertainty than the unpruned model. This is also reflected in their distributions of aleatoric uncertainty. The epistemic uncertainty narrows slightly for both the pruning methods. Pruning also affects the uncertainty calibration error, see Table~\ref{tab:calibration}. We find that while cUCE of mutual information does not change significantly, the predictive entropy and average entropy values have a higher calibration error, the effect being worse for SNR pruning. 

\begin{table}
\centering
\caption[Uncertainty Calibration Error]{Percentage classwise Uncertainty Calibration Error (cUCE) on MiraBest Confident test set for our BBB-CNN model trained with a Laplace prior. Results are shown for the unpruned model and the model pruned to its threshold limit for SNR and FIM-based pruning. The percentage cUCE is shown separately for the predictive entropy (PE), mutual information (MI) and average entropy (AE) as calculated on the MiraBest Confident test set.}
	\begin{tabular}{llccc}
		 &  & \multicolumn{3}{c}{\textbf{\% cUCE}} \\\cmidrule{3-5}
    \textbf{Prior} & \textbf{Pruning}  & \textbf{PE} & \textbf{MI}& \textbf{AE} \\
    \hline
    Laplace & Unpruned & 9.69  &  16.37 & 10.84\\
        & SNR & 14.35 & 16.82 & 13.93  \\
        & FIM & 13.43 & 15.29 & 11.25\\ 
   	\end{tabular}
   \label{tab:calibration}
\end{table}



\section{Conclusions}
\label{sec:conclusion}

In this work we find that whilst Bayesian neural networks using variational inference are useful for returning the degree of model confidence in an individual classification, and that on average this seems to agree with more general degrees of confidence assigned by human classifiers, there is still work that needs to be done on the development of these models. Most specifically, the effect of overly curated training data samples requires additional investigation and we suggest that a key element for exploring this problem in future may be the availability of radio astronomy training sets that do not only present average or consensus target labels, but instead include all individual labels from human classifiers.

Code for this work is available at: \url{https://github.com/devinamhn/RadioGalaxies-BBB}.

\begin{ack}
AMS gratefully acknowledges support from an Alan Turing Institute AI Fellowship EP/V030302/1.
\end{ack}

\clearpage
\bibliographystyle{unsrt85}
\bibliography{references}

\clearpage

\end{document}